\def\draftversion{false}
\RequirePackage{ifthen}
\ifthenelse{\equal{\draftversion}{true}}{
  \documentclass[aps,prb,10pt,galley,amsmath,amssymb,
                 floatfix,
                 citeautoscript,superscriptaddress]{revtex4}
}{
  \documentclass[aps,prb,10pt,twocolumn,amsmath,amssymb,
                 citeautoscript,superscriptaddress]{revtex4-1}
}

\usepackage{graphicx}% Include figure files
\usepackage[usenames]{color} % colors
\usepackage{bm} % bold math
\usepackage[update,prepend]{epstopdf}
\usepackage{soul} % for \st
\ifthenelse{\equal{\draftversion}{true}}{
  \marginparwidth 2.7in
  \marginparsep 0.5in
  \newcounter{comm} % counter for commentaries
  \def\commnext{\stepcounter{comm}}
  \def\commtext{{\bf\color{blue}[\arabic{comm}]}}
  \def\commmar{{\bf\color{blue}[\arabic{comm}]}}
  \def\dvm#1{\commnext\marginpar{\small DV\commmar: #1}\commtext}
  \def\jkm#1{\commnext\marginpar{\small JK\commmar: #1}\commtext}
  \def\hkm#1{\commnext\marginpar{\small HK\commmar: #1}\commtext}
  \def\mlab#1{\marginpar{\small\bf #1}}
  
}{
  \def\dvm#1{}
  \def\jkm#1{}
  \def\hkm#1{}
  \def\mlab#1{}
  
}
\def\beq{\begin{equation}}
\def\eeq{\end{equation}}

\def\z2{$\mathbb{Z}_2$}

\def\ket#1{\vert#1\rangle}

\def\TC{T_{\rm C}}
\def\KP{{\bm k}{\cdot}{\bm p}}

\newcommand{\Mbar}{\mkern 3.5mu\overline{\mkern-3.5mu M\mkern-1.5mu}\mkern 1.5mu}
\begin{document}

%===========================%
% TITLE PAGE                %
%===========================%

\title{Nearly triple nodal point topological phase in half-metallic GdN}

\author{Jinwoong Kim}
\author{Heung-Sik Kim}
\author{David Vanderbilt}
\email{dhv@rutgers.edu}
\affiliation{
Department of Physics \& Astronomy, Rutgers University,
Piscataway, New Jersey 08854, USA}

\date{\today}
\begin{abstract}
Recent developments in topological semimetals open a way to realize
relativistic dispersions in
condensed matter systems. One recently studied type of topological feature
is the
``triple nodal point'' where three bands become degenerate. In contrast to Weyl and
Dirac nodes, triple nodal points, which are protected by a rotational symmetry,
have nodal lines attached, so that a characterization in terms of a
chirality is not possible. Previous studies of triple nodal points considered
nonmagnetic systems, although an artificial Zeeman splitting was used to
probe the topological nature. Here instead we treat a ferromagnetic
material, half-metallic GdN, where the splitting of the triple nodal points
comes from the spin-orbit coupling. The size of the splitting ranges from
15 to 150\,meV depending on the magnetization orientation, enabling a
transition between a Weyl-point phase and a ``nearly triple nodal point'' phase
that exhibits very similar surface spectra and transport properties
compared to a true triple-node system. The rich topological surface
states, manipulable via the orientation of the magnetization, make
half-metallic GdN a promising platform for future investigations and
applications.

\end{abstract}
%\pacs{}% PACS

\maketitle

%===========================%
% MAIN TEXT                 %
%===========================%

%=================================================
\section{Introduction}
%=================================================

In the last decade, an enormous expansion in studies of topological
materials has opened a powerful new perspective in materials science
\cite{TI_Hasan,TI_SCZhang,TCI_LFu,QAHE_SCZhang,Bradlyn2017}.
While topological insulators are classified by integer
Chern numbers or $Z_2$ indices
\cite{TI_Hasan,TI_SCZhang,TCI_LFu,QAHE_SCZhang}, topological
semimetals\cite{TSM_ZFang,TSM_SRyu,TSM_Bansil} may be characterized
by the type of low-energy excitations they admit, in analogy with the
description of elementary particles in high-energy physics. In
particular, the excitations near a Dirac or Weyl point in a topological
semimetal behave similarly to the massless Dirac and Weyl fermions
that arise in the quantum field theory of elementary particls.
A Dirac point corresponds to a point fourfold degeneracy resulting from a
crossing of two-fold degenerate bands in momentum space; since the Berry flux
surrounding such a point vanishes, it has no net chirality. By
contrast, a Weyl point results from a crossing of just two bands,
and depending on its chirality, either emits or absorbs a
$2\pi$ quantum of Berry flux. As a consequence, Fermi arc states emerge
in the surface Brillouin zone (BZ) connecting the projected locations
of the Weyl points.

A recently studied three-fold band crossing point, referred to as a triple
nodal point (TNP), is protected by a crystalline (typically $C_3$ rotation)
symmetry\cite{Heikkila2015,Heikkila2016,Bradlynaaf5037,
XDai2016,Soluyanov2016,Chang2017,BYan2017}.
In contrast to the case of Dirac and Weyl points, the Berry phase of
the TNP is ill-defined due to the inevitable presence of nodal lines
attached to the TNP, which prohibits the occurrence of a gapped surface
enclosing a single TNP.
Although the formation of surface states has been demonstrated for
several TNP materials, the identification of a general
feature expected in the surface states, analogous to the Fermi arc
states, has remained elusive.
Moreover, the surface-state features are likely to be obscured if more
than one TNP projects to the same point on the cleavage surface; this
commonly occurs if that surface is orthogonal to the primary
rotation axis, which is the case in most of the suggested TNP
metals proposed to date\cite{Heikkila2015,
Heikkila2016,Bradlynaaf5037,XDai2016,Soluyanov2016,Chang2017}.

Here, we focus instead on half-metallic GdN possessing three perpendicular $C_4$
rotational axes so that at least two pairs of TNPs are exposed on a surface.
%\paragraph{Half-metallic GdN}
GdN and most of the rare earth monopnictide compounds occur in the rocksalt
structure and exhibit a variety of magnetic and transport
properties\cite{Duan2007,Natali2013}.
Early systematic theoretical studies\cite{Aerts2004,Horne2004,Szotek2004}
on the rare earth nitrides found a range of electronic structures
from narrow gap insulators (TbN, DyN, HoN) to half-metallic ferromagnets
(PrN, NdN, PmN, SmN, EuN, GdN) and ordinary metallic ferromagnetic materials
(CeN, ErN, TmN, YbN). Among the half-metallic ferromagnets, GdN
exhibits the highest Curie temperature ($\TC$)
of 58\,K\cite{GdN_Li}
and is reported to be a Chern insulator in an ultrathin two-dimensional
layer form \cite{ZLi},
suggesting potential Weyl nodes might emerge in three dimensions
BZ\cite{GXu2011}.
Although its exact band gap is still under debate even after intense
study\cite{Wachter1980,Aerts2004,Duan2005,Lambrecht2000,NMR_Leuenberger2005,Granville,Trodahl2007,Wachter2012},
there have been consistent reports that the band
gap decreases upon magnetic ordering
below $\TC$\cite{Trodahl2007,Yoshitomi2011}, external
pressure\cite{Duan2005,Yoshitomi,Kagawa2014}, and external magnetic
field\cite{Lambrecht2000}.

In view of the similarity of the electronic structure and tunability
of the band gaps in rare earth monopnictides, we have chosen to focus
here on GdN as a representative material for in-depth study.
We find that GdN exhibits a ``nearly triple nodal point'' (NTNP)
topological phase, analogous to the TNP phase but with a very small lifting
of the degeneracy of the TNPs.
The NTNPs come in pairs centered
on the three $X$ points in the BZ.  Because the spin-orbit coupling (SOC)
is so weak on N, while the spin splitting of the Gd orbitals is so large 
that the Gd SOC is largely quenched,
the SOC-induced splitting at each of the NTNP is quite small.
In fact, the system is characterized by the presence of several distinct
energy scales, with the
hopping-controlled band width dominating the exchange splitting which
in turn is much larger than the SOC, leading to a complex electronic
structure.

Because the splitting of the TNP is so weak,
the NTNP phase is found to have qualitatively similar surface spectra and
transport properties compared to a true TNP phase. Interestingly,
depending on the magnetic moment orientation, some of the
NTNPs decompose into conventional Weyl points.  As a result, we predict
that a selected rotation of the magnetization with external field can drive
transitions between Weyl and NTNP behavior in selected nodal regions,
with associated transitions in the surface-state topology.

%=================================================
\section{Methodology}
%=================================================

In order to investigate the electronic properties of GdN, \emph{ab initio}
calculations are carried out using {\small VASP} and wannierized using
the {\small{VASP-WANNIER90}} interface\cite{Kresse96a,Kresse96b,Mostofi}
to arrive at a tight-binding description of first-principles quality.
The pseudopotential is of the projector-augmented-wave type
\cite{Blochl94} as implemented in {\small{VASP}}\cite{KressePAW}.
The generalized gradient approximation exchange-correlation functional
is employed as parameterized by Perdew, Burke and Ernzerhof\cite{PBE} with
a Coulomb $U$ of 4.5\,eV on the Gd $f$ orbits.
The plane wave basis is expanded up
to the cutoff energy of 400\,eV and 12$\times$12$\times$12 $k$-mesh grid
is used in the {\small{VASP}} calculations. Six N~$p$ and ten Gd~$d$
atomic orbitals are projected for the wannier representation with spin
polarization but without SOC. The atomic SOC is then included in the
tight-binding Hamiltonian. The surface states for a semi-infinite
geometry are calculated
by employing an iterative surface Green's function method\cite{Lopez_SGF,KHiori_SGF}.
Landau level spectra are calculated using a symmetry-constraint $\KP$
model with Peierls substitution\cite{Rhim2015,Chang_LL}.

%=================================================
\section{Results and Discussion}
%=================================================

%-------------------------------------------------
\subsection{Without spin-orbit coupling}
%-------------------------------------------------

\begin{figure*} [t]
\includegraphics[width=8.6cm]{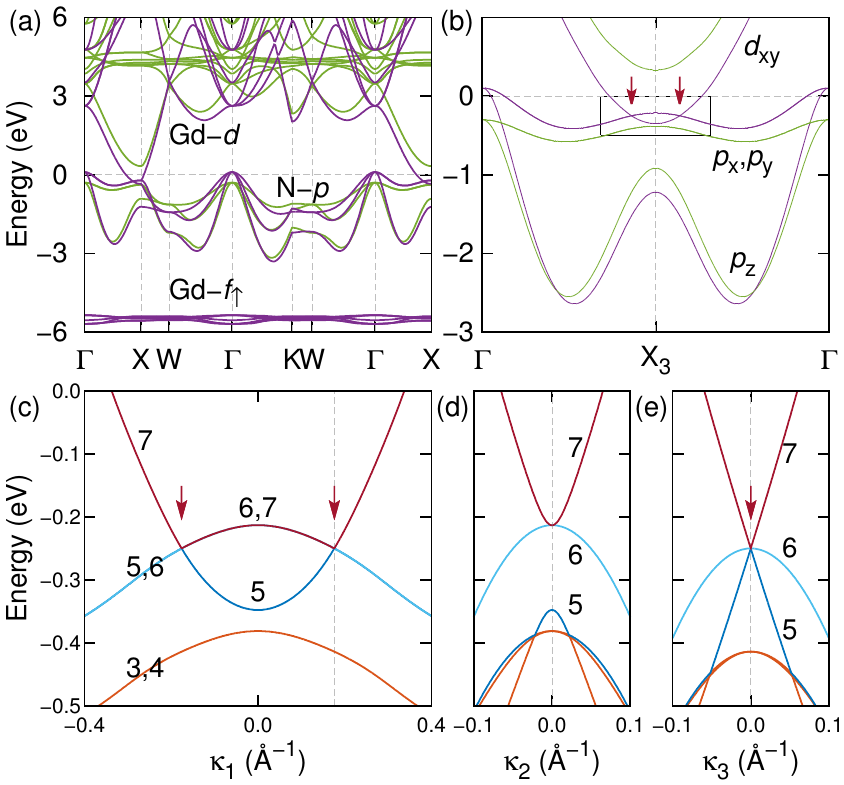}
\includegraphics[width=9.2cm]{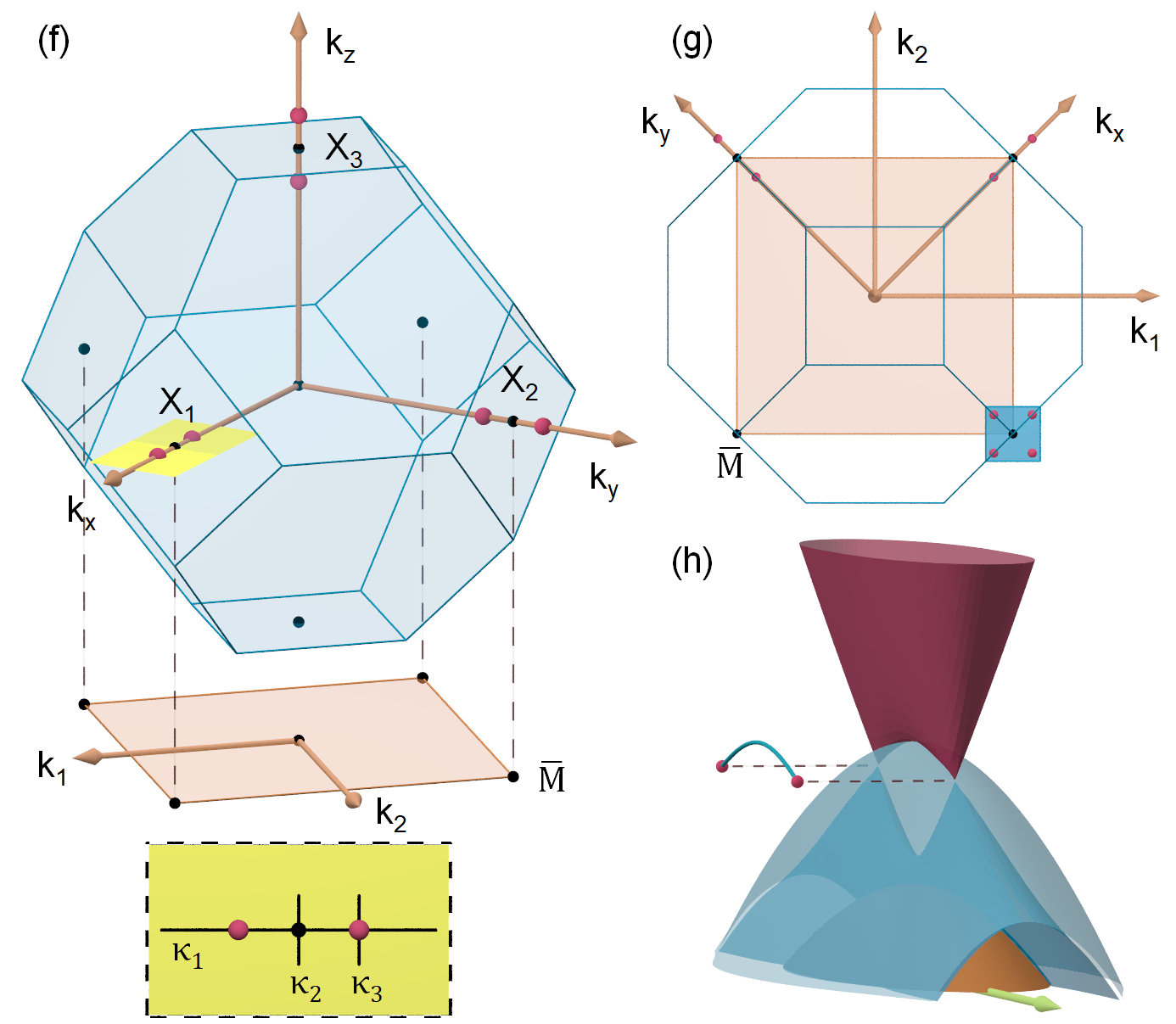}
\caption{(a-e) Band structure of GdN calculated without SOC. In (a,b)
the purple (green) colors represent spin majority (minority) bands,
while in the blow-ups in (c-e) the colors are keyed to indicated band
index numbers, where band 6 is the highest valence band.
Triple nodal points are marked with red arrows. (f) Relation of
the bulk Brillouin zone (BZ) to the two-dimensional (2D) projected
BZ on the (001) surface.
Six red dots denote triple nodal points; yellow rectangle defines the three
variables $\kappa_1$, $\kappa_2$, and $\kappa_3$ used as the horizontal
axes in (c-e). (g) Top view of
the BZ, showing the overlap between bulk and surface BZs.
Blue square is the area used for constant-energy scans in 
Figs.~2 and 4-6.
(h) 2D band structure plotted on the $k_{z}\!=\!0$ plane near the $X_1$ point;
the arrow at the base is directed toward the $\Gamma$ point, and the
shifted blue line terminating in two dots illustrates
the nodal line connecting the two triple nodal points.}
\label{fig:1bulk_wo_soc}
\end{figure*}

%\paragraph{Type-A triple nodal points without SOC}

The electronic structure of GdN in the absence of SOC is shown in
Fig.~\ref{fig:1bulk_wo_soc}. Near the Fermi level, the valence and
conduction bands mostly consist of N~$p$ and Gd~$t_{2g}$ orbitals respectively.
In the nonmagnetic phase, a small indirect gap appears between the $\Gamma$
and $X$ points.
When cooled down below the Curie temperature of 58\,K\cite{GdN_Li}, GdN
becomes half-metallic due to the opposite sign of the Zeeman splitting
on Gd and N atoms,
causing a band inversion only in the majority-spin channel. The band
inversion does not open a mass gap because the bands belong to different
irreps of the $C_4$ rotations about the primary $\Gamma$-$X$
axes. For example, concerning $C_4$ rotations about the $z$ axis,
the states on the $\Gamma$-$X_{3}$ line obey
$R\left(C_{4}^{\hat{z}}\right)\ket{d_{xy}}=-1\ket{d_{xy}}$ and
$R\left(C_{4}^{\hat{z}}\right)\ket{p_{x}\pm ip_{y}}=\pm i\ket{p_{x}\pm
ip_{y}}$. Note that the $\ket{p_{x}\pm ip_{y}}$ valence-band states
remain doubly degenerate, since in the absence of SOC the orbital moment
does not couple with the spin moment. The crossing point is thus triply
degenerate and is referred to as a TNP\cite{XDai2016,Soluyanov2016,
BYan2017,Zhang2017,Chang2017}.

Figures~\ref{fig:1bulk_wo_soc}(c) and (h) show that each pair of
TNPs near an $X$ point is connected by a nodal line. Since a cross
section of the nodal line is a quadratic touching point
[Fig.~\ref{fig:1bulk_wo_soc}(d)], the Berry phase
around the nodal line is zero
and no surface state is induced by the nodal line.
\cite{Soluyanov2016,Chang2017}
In the notation of Refs.~[\onlinecite{Soluyanov2016,Chang2017}],
this corresponds to a type-A TNP.
(Their type-B TNP is connected by several nodal lines 
lying off the symmetry axis in addition to the one lying on the axis.)
In contrast to the typical TNP materials, the TNPs of GdN are 
protected by $C_4$ rotational symmetry in the absence of SOC.%
\footnote{Note that $C_4$ symmetry cannot protect TNPs in the
   presence of SOC.  If time reversal is absent, there are only
   one-dimensional irreps, and their crossings generate simple
   Weyl points.  If it is present, irreps $\theta=\pm\pi/4$ and
   $\pm3\pi/4$ come in time-reversal pairs, enforcing every
   crossing on the $C_4$ axis to be four-fold degenerate.  See also
   Ref.~[\onlinecite{Zhang2017}].}

\begin{figure} [t]
\includegraphics[width=8.6cm]{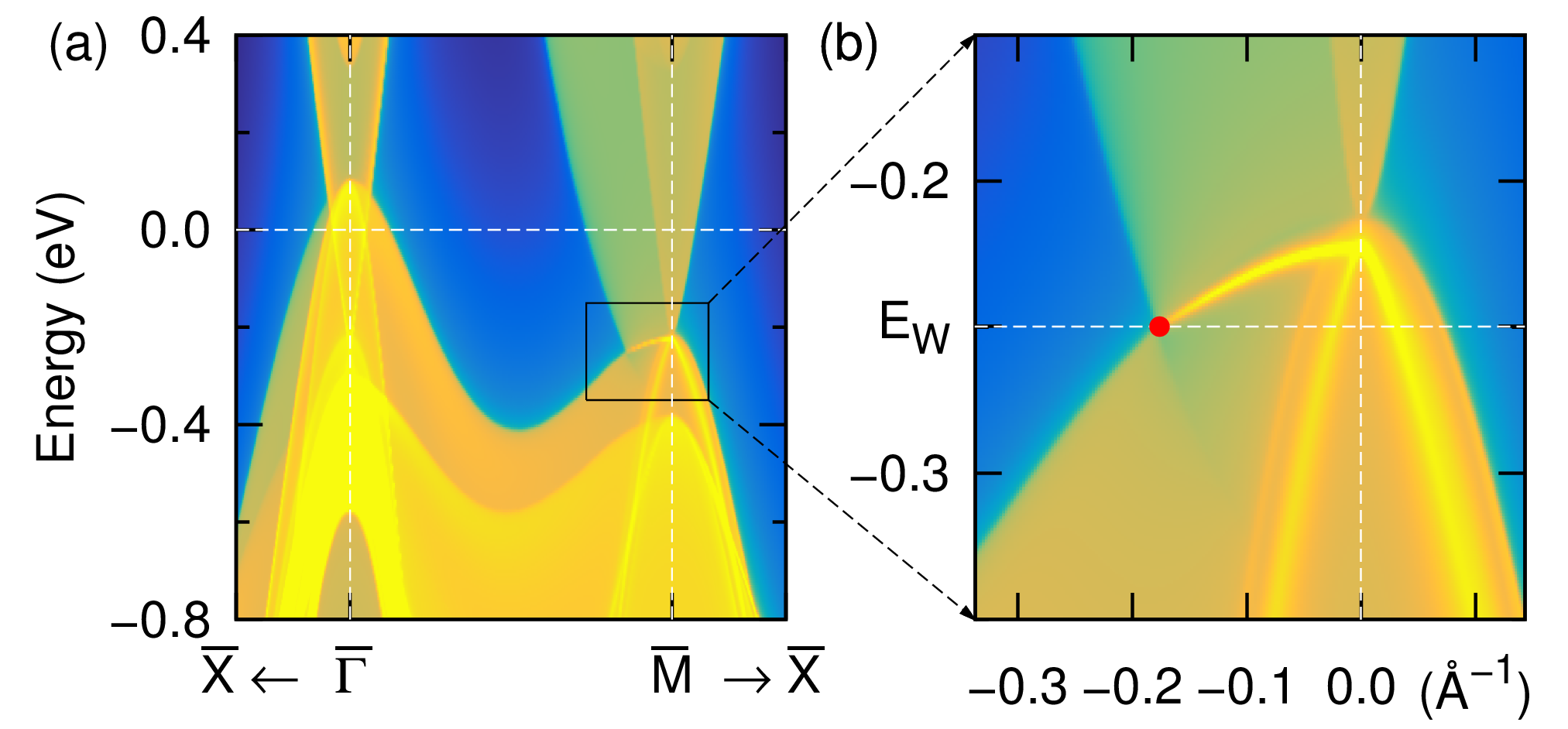}
\includegraphics[width=8.6cm]{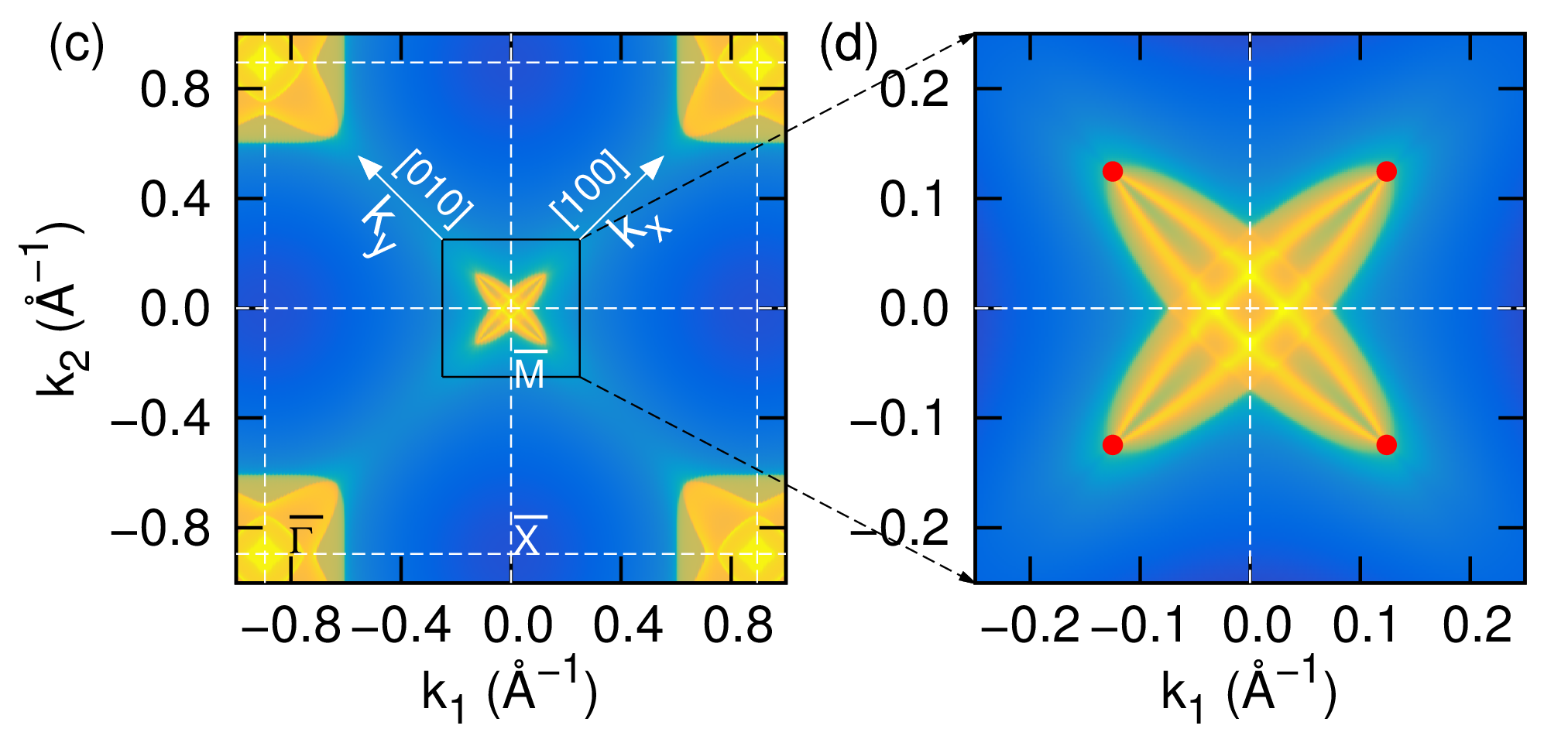}
\caption{Surface states on the (001) surface calculated for a
semi-infinite geometry without SOC. (a-b) Surface band
spectral function. (c-d) Constant-energy scan at $E=E_W$.
Bright (dim) yellow color represents surface (bulk) state. The
origin has been shifted to the $\Mbar$ point.  (b) and (d)
are zoomed in around the $\Mbar$ point; red dots indicate triple
nodal points.}
\label{fig:2surf_wo_soc}
\end{figure}

Unlike Weyl points and nodal loops, which generate
Fermi arcs and drumhead states respectively, previous work has not
identified a corresponding general feature expected in the surface-state spectrum of a TNP material.
In comparison with previously
reported TNP materials, the TNPs of GdN are sufficiently well
isolated from irrelevant bands that the resulting surface states
can be well characterized.

Figure~\ref{fig:2surf_wo_soc} shows the surface states of semi-infinite
GdN in the absence of SOC. The right panels are blow-ups of the left panels,
with the TNPs shown as red dots. Panels (c-d) are constant-energy
intensity plots on a plane containing four TNPs,
at the energy of the TNPs.
The surface states attributed to the TNPs appear
bright yellow in Figs.~2(b) and (d), in comparison to the dim bulk states
in dark yellow.
In panel (d),
the projected TNPs are clearly seen to be attached
by two branches of Fermi arcs.
Overall, the surface-state structure looks like two overlapping copies
of an elliptical dome rotated by
$90^{\circ}$ with respect to each other.
The shape of the surface-state structure is discussed and further
illustrated in Sec.~III-C. 
These elliptical domes, which have open sides below the two TNPs,
are detached from the conduction band, contrary to the case of the surface
states of a conventional Weyl phase.
The dome shape can be understood as a hybridization of two surface states
individually induced by two pairs of conventional Weyl points, as will
be discussed in more detail below.

%-------------------------------------------------
\subsection{With spin-orbit coupling}
%-------------------------------------------------

\begin{figure} [t]
\includegraphics[width=8.6cm]{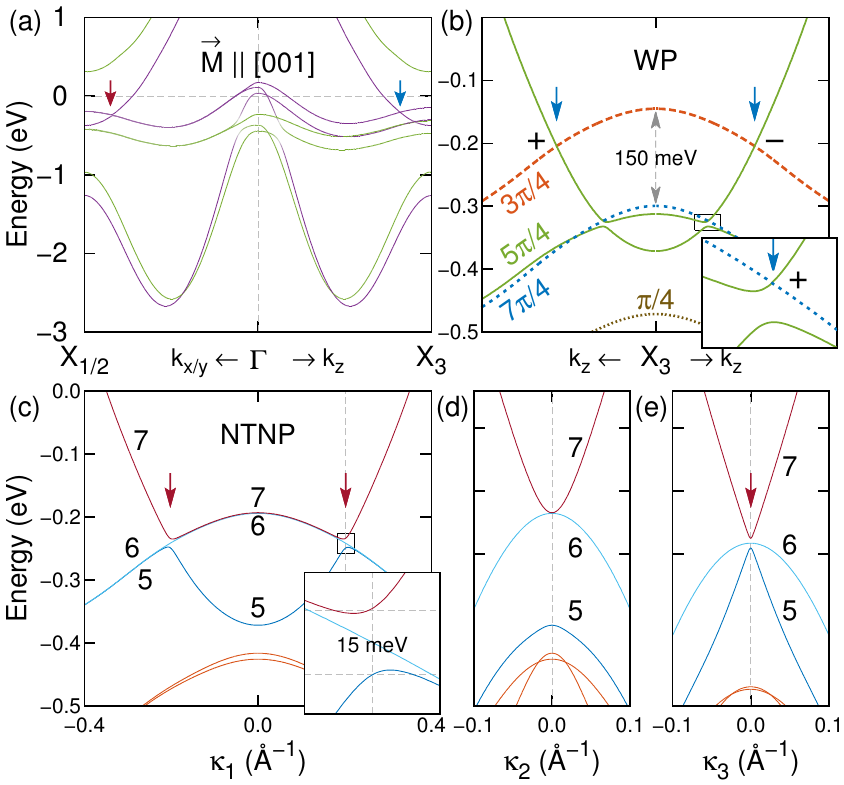}
\includegraphics[height=4cm]{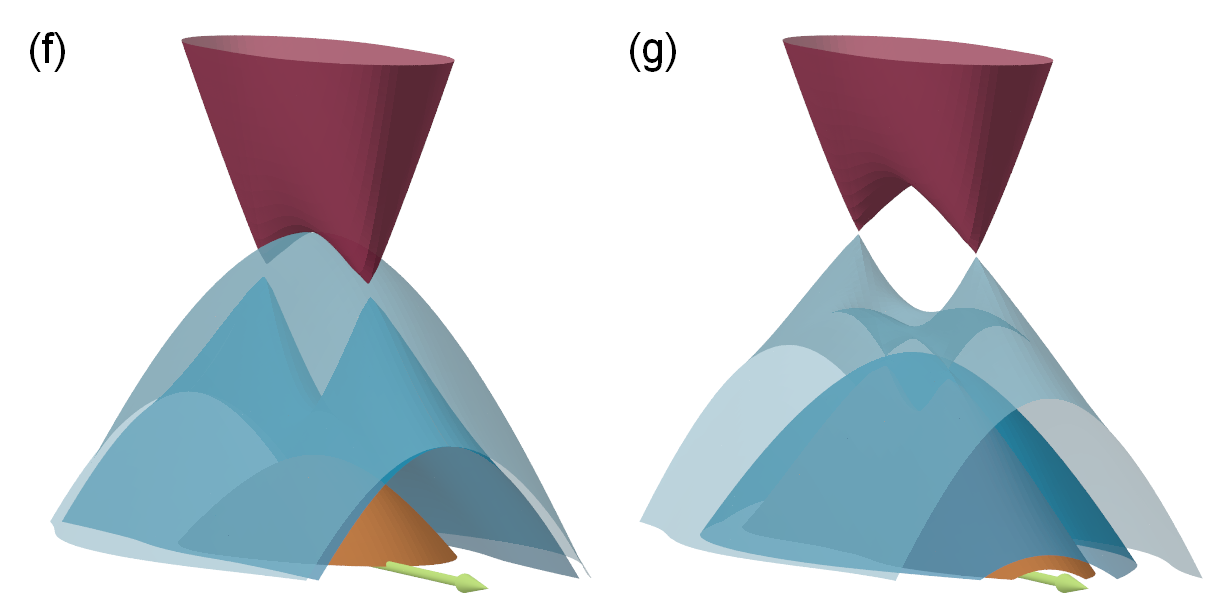}
\caption{Electronic structure of GdN calculated with SOC. The magnetic
moment is aligned along the [001] direction. (a) Band dispersion along the
three cartesian axes. The purple (green) colors represent spin majority 
(minority) bands. (b) Band structure on the $k_z$
axis, where band labels denote eigenvalues of the $C_4^{\hat{z}}$ operator.
(c-e) Band structure on $\kappa_1$, $\kappa_2$, and $\kappa_3$ lines.
Blue and red arrows denote Weyl and nearly triple nodal points, respectively.
(f) and (g) 2D band structure on $k_x$-$k_y$ and $k_z$-$k_x$
planes at $X_1$ and $X_3$, respectively. 
%Arrows at the bottom point toward the $\Gamma$ point.
The arrows at the base are directed toward the $\Gamma$ point.
}
\label{fig:3bulk_001}
\end{figure}

%\paragraph{Nearly triple nodal points}

%=========================================
% Nearly triple nodal points
When taking SOC into account, a direction of the net magnetic moment
has to be specified. If we take it along the $[001]$ direction,
the degeneracy of the $\ket{p_{x}\pm ip_{y}}$ valence bands
on the $k_z$ axis is significantly lifted,
whereas the corresponding states on the $k_x$ and $k_y$ axes are
hardly altered, as
shown in Fig.~\ref{fig:3bulk_001}. On the $k_x$ axis, for instance,
the band of $\ket{p_x}$ character is more dispersive and located lower in energy
than the $\ket{p_y}$ and $\ket{p_z}$ bands, due to stronger orbital
overlaps along the $\hat{x}$ direction. The large energy separation of
$\ket{p_x}$ and $\ket{p_y}$ bands
causes a weak coupling with the magnetic moment, ${\bm M}\parallel[001]$. As
will be discussed below, the resulting surface states are quantitatively
similar to those of the true TNP phase, as manifested by
turning off the SOC.
Therefore, we refer to the nearly triply degenerate crossing
points as ``nearly triple nodal points'' (NTNPs). 
Figures~\ref{fig:3bulk_001}(c-f)
show the zoomed-in band dispersion of the NTNPs corresponding to
Figs.~\ref{fig:1bulk_wo_soc}(c-e,h) respectively. The NTNPs
have a small gap opening of $\sim$15\,meV.

%=========================================
% Conventional Weyl point on the axis || M
When SOC is introduced, the $C_4$ rotational symmetries are generally
broken in the presence of a magnetic moment that is not parallel to
the rotation axis.
However, with the parallel magnetic moment ${\bm M}\parallel[001]$,
$C_4^{\hat{z}}$ symmetry still remains and Bloch states on the
$k_z$ axis are classified with the eigenvalues of the rotational operator,
$R\left(C_4^{\hat{z}}\right)$.
Figure~\ref{fig:3bulk_001}(b) shows the band structure labeled by
the phase
$\theta=\left\{\frac{\pi}{4},\frac{3\pi}{4},\frac{5\pi}{4},\frac{7\pi}{4}\right\}$
of the eigenvalues of the $C_{4}^{\hat{z}}$ operator.
The two bands having the same phase $\theta = 5\pi/4$
mix with each other and open a mass gap,
whereas others with distinct phases cross each other without a gap
opening, producing a conventional two-fold Weyl node.
The Chern number of a Weyl point can be determined by the phase difference
of the two crossing bands\cite{Fang_Bernevig2012,Tsirkin2017}. 
For example, the point at which a $\theta={3\pi}/{4}$ band 
crosses up (down) through a $\theta={5\pi}/{4}$ band with increasing $z$ 
has a Chern number of $+1$ ($-1$),
and would serve as the terminus for a single Fermi arc on
the surface. The other crossing points between $\theta={5\pi}/{4}$ and
$\theta={7\pi}/{4}$ bands produce another 
pair of Weyl points, of which one is shown in the inset with 
a Chern number of $+1$. 
The 2D band structure in Fig.~\ref{fig:3bulk_001}(g) shows that the 
parabolic band is shifted down in energy due to the Zeeman splitting,
leaving two conventional Weyl nodes prominently exposed.
Since the Weyl points are robust unless they are mutually
annihilated, the Weyl points still survive under a small rotation of the
magnetic moments even without the $C_{4}^{\hat{z}}$ symmetry.
Under the rotation, the Weyl points are found to migrate
in the vicinity of the primary axis (not shown here).

%-------------------------------------------------
\subsection{Manipulation via magnetization and strain}
%-------------------------------------------------

%=========================================
% Manipulation of the effective degeneracy
%
It is important to note that the NTNPs appear on the axes perpendicular
to the magnetic-moment direction, while the Weyl points
lie on the parallel axis. Thus, if one rotates the magnetic moment
from $\bm{M}\parallel[001]$ toward the [100] direction (keeping
$M_y=0$),
the NTNPs on the $k_y$ axis are unaffected, whereas 
those on the $k_x$ axis split into Weyl points because the 
degeneracy of the $\ket{p_{y}\pm ip_{z}}$ valence-band states 
is lifted by the finite $M_x$.

\begin{figure}
\begin{center}
\includegraphics[width=8.6cm]{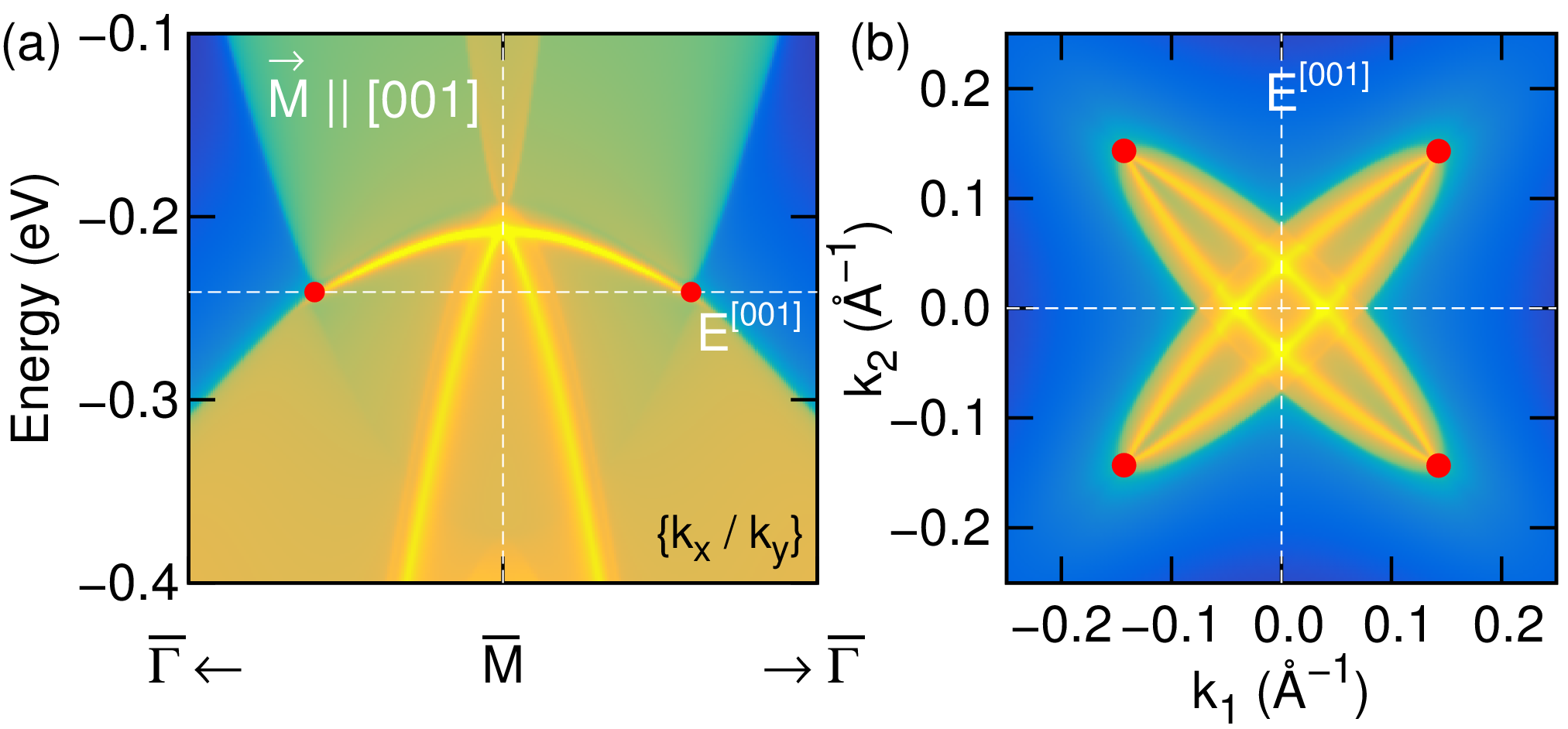}
\vspace{-0.8cm}
\end{center}
\includegraphics[width=8.6cm]{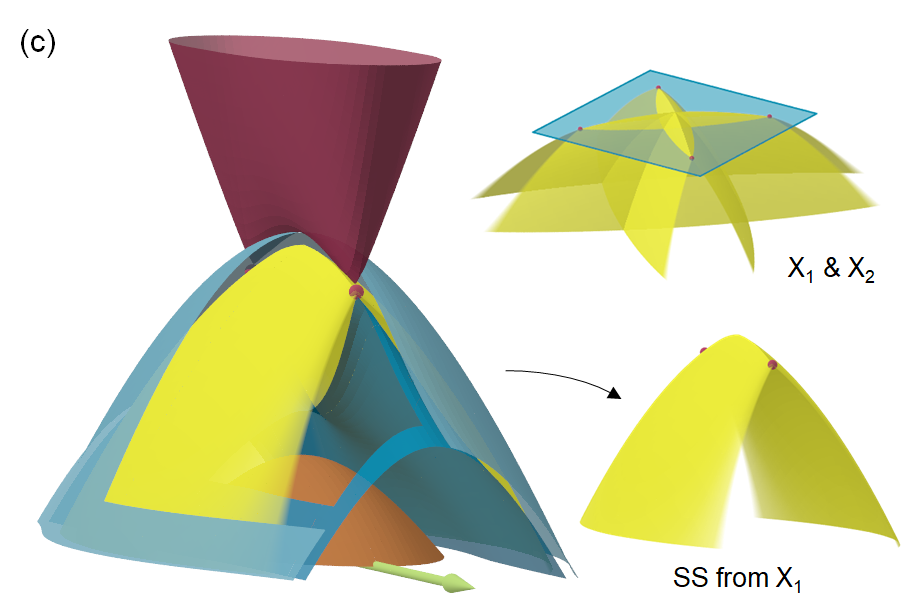}
\caption{Electronic structure on the (001) surface calculated for 
a semi-infinite geometry with the magnetization orientation of [001] 
direction. (a) Band structure at the
$\Mbar$ point. (b) Constant-energy scan at $E=E^{[001]}$. Bright yellow
color denotes intense spectral density. (c) Schematic view of the nearly
triple-nodal-point
surface state emerging from the $X_1$ and $X_2$ points. Red dots are nearly
triple nodal points.}
\label{fig:4surf_001}
\end{figure}

\begin{figure}[t]
\includegraphics[width=8.6cm]{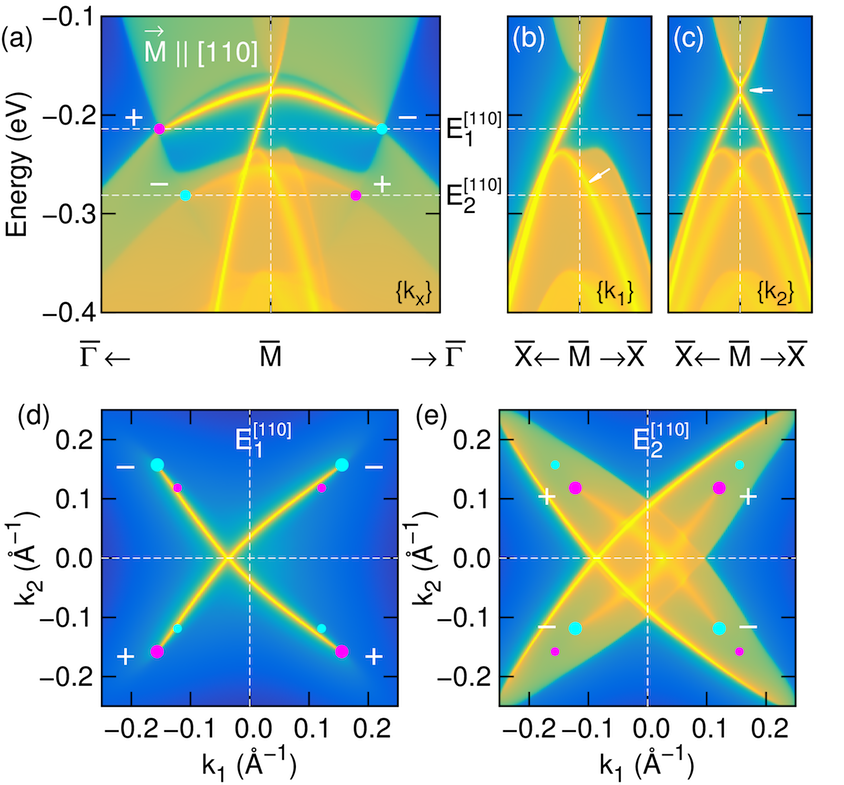}
\caption{Electronic structure on the (001) surface calculated for a 
semi-infinite geometry with the
magnetization along the [110] direction. (a-c) Band structure at the
$\Mbar$ point along $k_x$, $k_1$, and $k_2$ directions
[see Fig.~\ref{fig:1bulk_wo_soc}(g)].
Constant-energy-scans at 
(d) $E=E_1^{[110]}$ and (e) $E=E_2^{[110]}$. 
Magenta (cyan) dots denote the position of Weyl points
with positive (negative) chirality. In (d-e),
large dots lie on the energy
plane of the plot, while small dots lie off the plane. }
\label{fig:5surf_110}
\end{figure}

\begin{figure*}[t]
\includegraphics[width=17.8cm]{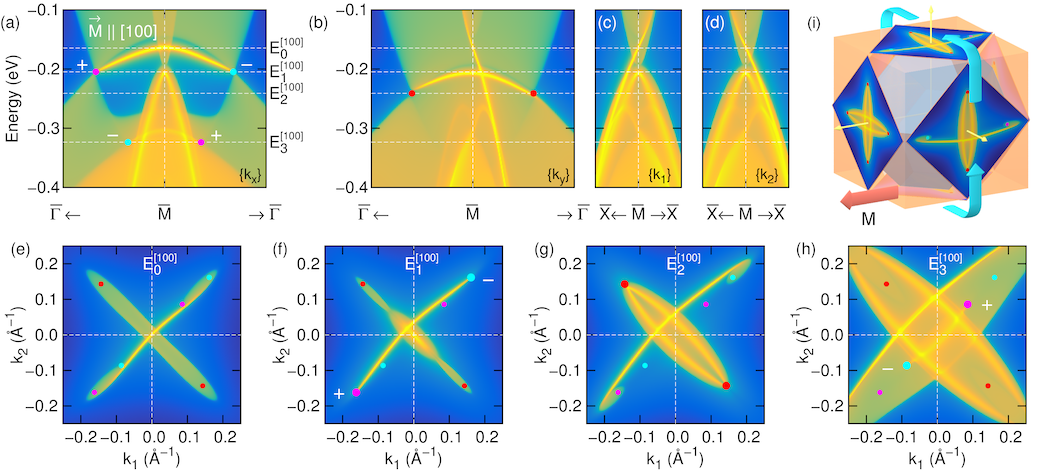}
\caption{Electronic structure of the (001) surface calculated for a 
semi-infinite geometry with the
magnetization along the [100] direction. (a-d) Band structure at
the $\Mbar$ point along $k_x$, $k_y$, $k_1$, and $k_2$ directions. 
Constant-energy-scans at (e) $E=E_{0}^{[100]}$,
(f) $E=E_{1}^{[100]}$, (g) $E=E_{2}^{[100]}$, and (h) $E=E_{3}^{[100]}$.
Red dots represent nearly triple nodal points. 
Magenta (cyan) dots denote Weyl points with
positive (negative) chirality. In (e-h),
large dots lie on the energy
plane of the plot, while small dots lie off the plane.
(i) Schematic view of the chiral surface states. Red and blue arrows
indicate direction of magnetization and surface group velocity, 
respectively.}
\label{fig:6surf_100}
\end{figure*}

%=========================================
% Surface states upon Magnetic moment direction
% [001] direction
Because the magnetocrystalline anisotropy of GdN is very small,%
\footnote{According to our calculations, the total energy is equal
	within numberical accuracy for magnetization along [110] or [111],
	and only about 0.05\,meV per formula unit higher for magnetization
	along [100].}
one can easily control the magnetization orientation by applying
an external magnetic field.
In cooperation with the SOC, the magnetic moment of GdN is thus a
tool that can be used to manipulate the Weyl nodes in energy and
momentum space, and thus the
surface states as well. 
The calculated semi-infinite (001) surface states
are shown in Fig.~\ref{fig:4surf_001} -~\ref{fig:6surf_100} with respect to
the magnetic moment orientation. For a magnetic moment normal to the
surface, two pairs of NTNPs are projected on the surface BZ
and connected by elliptical dome-like surface states as illustrated in
Fig.~\ref{fig:4surf_001}(c). This surface state has similar
features as that of the true TNP phase presented above in the absence of the
SOC [Fig.~\ref{fig:2surf_wo_soc}(b) and (d)].

% [110] direction
A magnetic moment along the [110] direction makes the four NTNPs split into four
pairs of conventional Weyl points (Fig.~\ref{fig:5surf_110}).
This is different from the splitting of a Dirac point into two
Weyl points, in that here one Weyl point is located one band index
higher than the other Weyl point. For instance, four Weyl
points are crossings of valence and conduction bands (at an energy level
of $E^{[110]}_{1}$) while the other four Weyl points are
crossings of two valence bands (at an energy level of $E^{[110]}_{2}$).
Figure~\ref{fig:5surf_110}(d) shows two Fermi arcs connecting two pairs of
Weyl points on the constant-energy plane $E=E_1^{[110]}$. Since the Weyl
points are crossings of valence and conduction bands, the surface states
are guaranteed to connect valence and conduction bands crossing
the energy gap at $E_1^{[110]}$ [Fig.~\ref{fig:5surf_110}(a-c)].
Surface states from the other four Weyl
points on $E=E_2^{[110]}$  are immersed in the bulk valence bands
as marked by an arrow in Fig.~\ref{fig:5surf_110}(b).
Nevertheless, the constant-energy plot [Fig.~\ref{fig:5surf_110}(e)]
still shows embedded Fermi arcs connecting the Weyl nodes.
An arrow in Fig. 5(c) shows a small gap at the crossing points, implying
a finite interaction between the two Fermi arcs.

% [100] direction
The last case we discuss here is when the magnetic moment is aligned along
the [100] direction, which lifts the NTNPs on the $k_x$ axis but not on the $k_y$ axis as
shown in Fig.~\ref{fig:6surf_100}(a) and (b), respectively. The Fermi arcs
at relevant energy levels are plotted in Fig.~\ref{fig:6surf_100}(e-h),
showing the coexistance of the NTNP and the conventional Weyl
point phase.
It is noteworthy that the Fermi arc is tangentially attached to the hole
or electron pockets enclosing the Weyl points, as is
clearly demonstrated in Fig.~\ref{fig:6surf_100}(e).
This is expected based on
the analysis of Haldane\cite{Haldane2014}, but
to our knowledge this has not previously been demonstrated using
\emph{ab initio} calculations.
If the energy is decreased so that it falls below the Weyl point,
the Fermi arc becomes attached to the other side of the hole pocket
[Fig.~\ref{fig:6surf_100}(g)], preserving the tangential attachment
in good agreement with the prediction.\cite{Haldane2014}

\begin{figure} [b]
\begin{center}
\includegraphics[width=8.6cm]{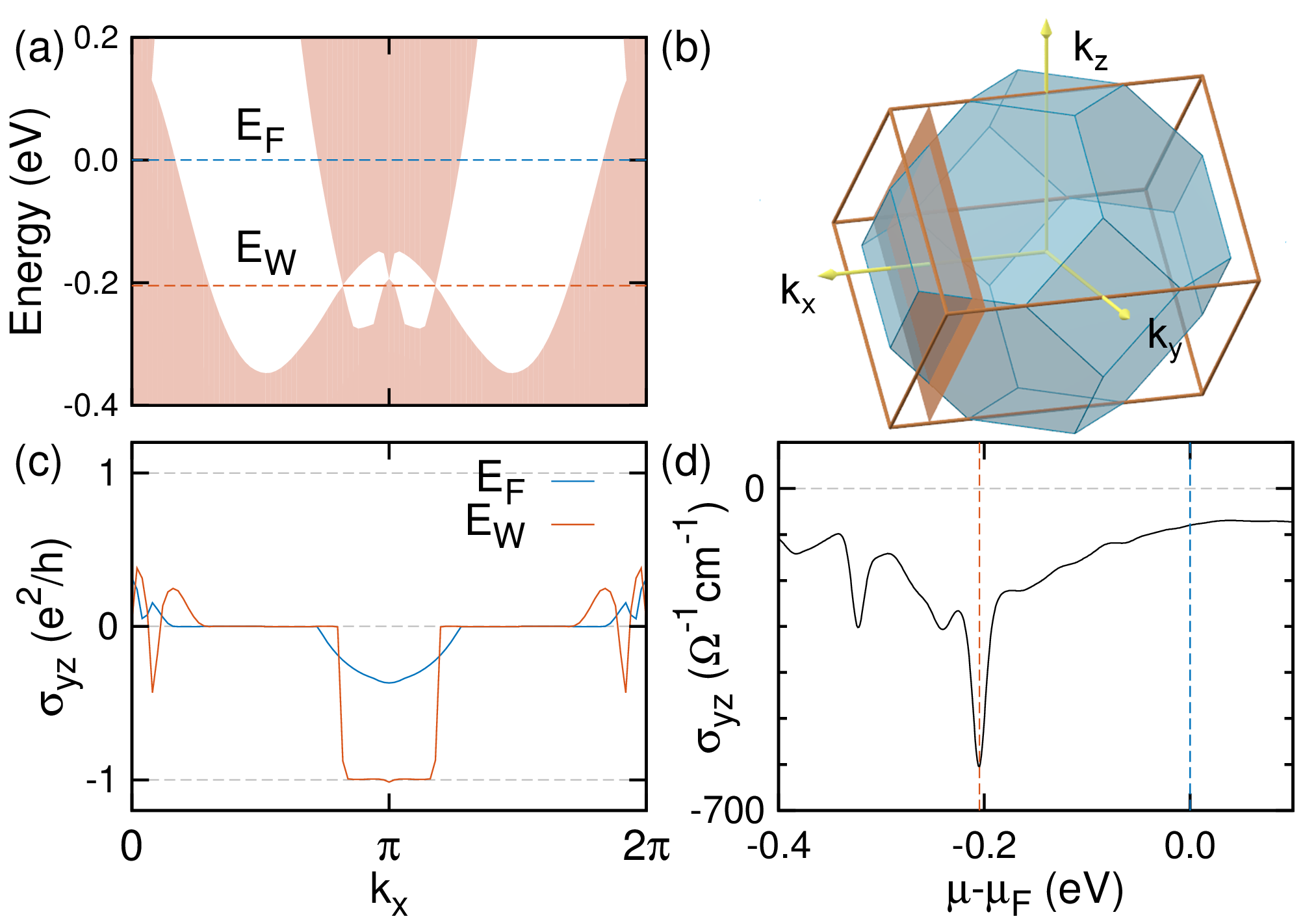}
\end{center}
\caption{Anomalous Hall conductivity for magnetization along the
[100] direction. (a) Electronic states projected on the $k_x$ axis 
from the tetragonal Brillouin zone (BZ) having the same volume as the
conventional Wigner-Seitz BZ, as shown in (b).
(c) Two-dimensional anomalous Hall conductivity vs.~$k_x$
calculated on constant-$k_x$ planes, one of which is shown as a colored
square in (b), including all states below $E_F$ (blue) or $E_W$ (brown).
(d) Bulk anomalous Hall conductivity as a function of chemical
potential.}
\label{fig:7AHC}
\end{figure}

% Chiral surface channel
Figure~\ref{fig:6surf_100}(i) illustrates the surface states of a
cubic crystallite of GdN, showing that
the surfaces parallel to the magentic moment direction have chiral conducting
channels associated with the bulk Weyl nodes. These chiral channels
circulate in a right-handed manner relative to the magnetic moment
direction. This is true not only for
${\bm M}\parallel[100]$, but also for an arbitrary direction of $\bm{M}$,
because the component of the Weyl-point chiral dipole moment
is proportional to the
magnetization in each Cartesian direction.
Figure~\ref{fig:7AHC} shows that in the simplest case
of ${\bm M}\parallel[100]$, only one pair of Weyl points lies on the $k_x$
axis in the vicinity of the Fermi level, giving non-zero anomalous Hall
conductivity $\sigma_{yz}$. Panel~\ref{fig:7AHC}(c)
shows the partial Chern numbers $\mathcal{Z}$ where
\begin{align}
\sigma_{yz}\left(k_{x}\right) = -\frac{e^{2}}{h}\mathcal{Z}\left(k_{x}\right),
\end{align}
calculated in 2D $(k_{y},k_{z})$ momentum space as a function of
$k_x$ for two chemical potentials. It shows plateaus in
gapped windows of $k_x$ where $\sigma_{yz}$ is well quantized to either
0 or $-1$, corresponding the region
between the two Weyl points associated with
the surface chiral channels. Unfortunately, the Weyl points are
separated
from the Fermi level by about $-0.2$\,eV, suggesting that gating or
doping would be
required to measure the chiral transport properties shown in
Fig.~\ref{fig:7AHC}(d).

\begin{figure} [t]
\begin{center}
\includegraphics[width=8.6cm]{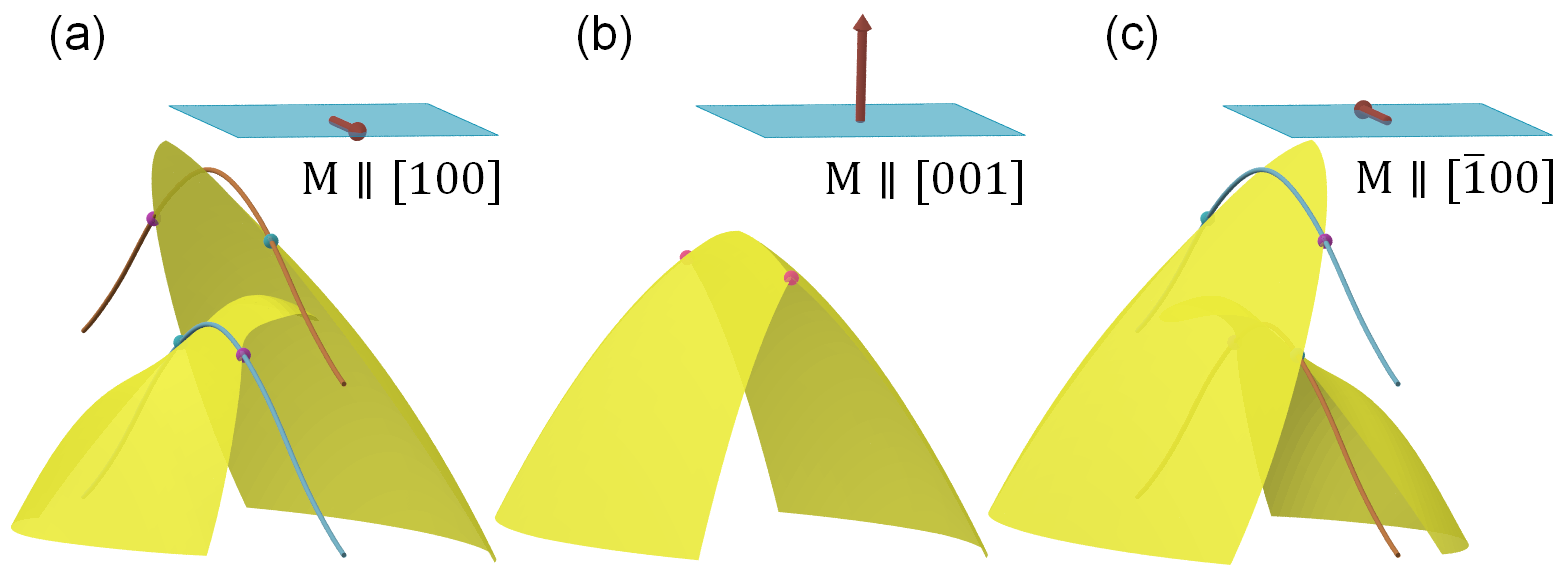}
\end{center}
\caption{Schematic view of the GdN surface-state transition 
on the (001) surface for magnetization 
orientation along (a) $[100]$, (b) $[001]$, and (c) $[\bar{1}00]$.
Orange and blue lines represent two bulk bands on the $k_x$ axis having
$C_4^{\hat{x}}$ eigenvalues of $\textrm{exp}\left(i3\pi/4\right)$ and
$\textrm{exp}\left(i7\pi/4\right)$, respectively [see Fig.~\ref{fig:3bulk_001}(b)]. }
\label{fig:8schematic}
\end{figure}

% transition of surface state
The dome-like surface state of the NTNP phase can be understood as a
transitional state between the two Weyl phases.  This is illustrated in
Fig.~\ref{fig:8schematic}, which
shows the topological surface state on the (001) surface 
induced by Weyl points or NTNPs on the $k_x$ axis.
When rotating the magnetic moment from the [100] to
the [001] direction, two initially separated valence bands become nearly
degenerate, inducing a pair of NTNPs. A further rotation to the [$\bar{1}00$]
direction splits the two valence bands again, in such a way that
the chirality is exchanged and the
surface states acquire the opposite group velocities.
Note that an external magnetic field will also split TNPs into
Weyl nodes,\cite{Soluyanov2016} and the rotation of the
applied field can cause a qualitatively similar transition in
the TNP surface states.
Thus, the TNP and NTNP phases act as intermediate neutral
states at which the chirality is reversed via the fusion and fission
of two chiral surface states.

\begin{figure} [t]
\includegraphics[width=8.6cm]{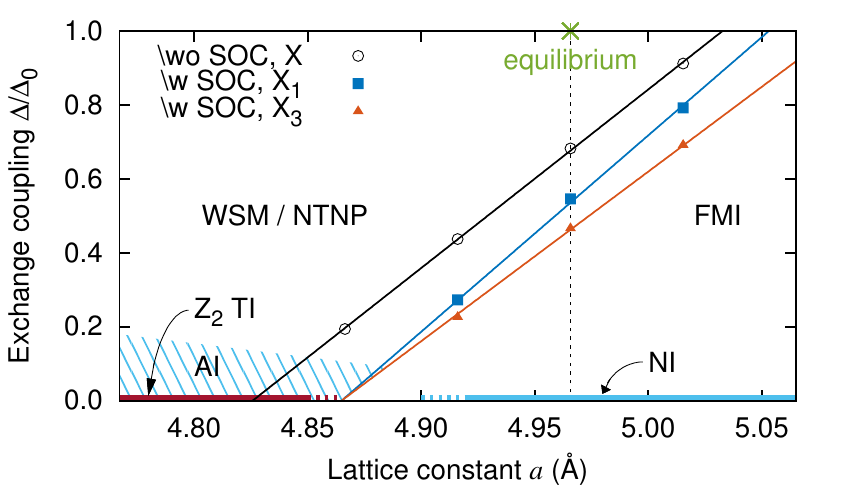}
\caption{Topological phase diagram of GdN as a function of strain and exchange
coupling with the  magnetic moment along [001]. Plotted lines indicate
closures of the direct band gap: without SOC (black), simultaneously
at all three $X$ points; and with SOC, at $X_1$ (blue) and $X_3$
(red).  Vertical dashed line indicates the equilibrium lattice
constant. Acronyms are WSM for Weyl semimetal, NTNP for nearly
triple nodal point, FMI for ferromagnetic insulator, TI for
topological insulator, AI for axion insulator, and NI for normal
insulator.}
\label{fig:9phase}
\end{figure}

An essential prerequisite for the occurrence of the band crossings that
we have analysed above is the presence of a band inversion at the
$X$ points, which is quite sensitive to the
lattice constant and to the exchange coupling strength.
Figure~\ref{fig:9phase} shows the calculated location of the
band gap closure as a function of these two parameters.  We find that
both compressive strain and stronger exchange coupling enhance the
band inversion, in agreement with previous reports
\cite{Trodahl2007,Yoshitomi2011,Duan2005,Yoshitomi,Kagawa2014,Lambrecht2000}.
The SOC also strengthens the band inversion, especially on the primary axis
parallel to the magnetic moment, due to its larger band splitting.
Therefore, in the region between the blue and red
lines, GdN has only one pair of Weyl points.
It is noteworthy
that a uniaxial pressure may widen the single-pair area by causing
compressive and tensile strains along the $X_3$ and $X_{1,2}$ axes respectively,
as will be discussed below.
Once all three band inversions have occurred,
the number of NTNP is determined by the magnetic
moment direction.
If the magnetic moment disappears at $T>\TC$,
the $C_4$-rotational and time-reversal symmetries are recovered 
together with the spin degeneracy of each band. On the Cartesian axes, 
the spin-degenerate valence and conduction bands near the Fermi level
have $C_4$ eigenvalues of $\theta=\pm3\pi/4$,
implying a mass gap where they cross.
The band inversion at the three $X$ points implies that the
topology of the bands lying below the global direct gap is that
of a $Z_2$ time-reversal topological insulator, and
if a weak ordering of the magnetic moments is turned on,
this turns into an axion insulating phase
in a narrow window of the exchange coupling parameter 
labeled as ``AI'' in Fig.~\ref{fig:9phase}.

%\paragraph{Anisotropic strain}
The fact that the three band inversions occur on three orthogonal
primary axes opens the possibility of tuning these gaps individually
via anisotropic strains.  A uniaxial compressive stress, for instance,
should enhance the band inversion on the primary axis while
reducing or eliminating it on the other two axes. Thus, this might
result in just a single band inversion, with one pair of NTNPs
or one pair of Weyl points (depending on magnetization direction)
on the primary axis.
A biaxial stress might induce two pairs of Weyl points without any
NTNPs. It is worth recalling that the effective
degeneracy on the $k_x$ axis originates from N $p_{y}$ and $p_{z}$ orbitals
and is lifted by distinct lattice constants $a_y \ne a_z$.
Under biaxial pressure, therefore, the NTNPs cannot emerge
on the two in-plane axes, regardless of the magnetic moment direction.

%-------------------------------------------------
\subsection{Landau level spectra}
%-------------------------------------------------

%\paragraph{Landau level spectra}
One characteristic feature of a TNP material is the appearance in
magnetotransport of equally spaced Landau levels crossing the Fermi 
level, instead of a single chiral level as in Weyl semimetals
\cite{Soluyanov2016,Chang2017}.
To investigate the magnetotransport properties of both 
TNP and NTNP phases, the Landau level spectra are calculated by 
performing a Peierls substitution in a $\KP$ model.
First, a simple $\KP$ model in the absence of SOC is constructed with a 
minimal basis set of $\ket{p_x}$, $\ket{p_y}$, and $\ket{d_{xy}}$ orbitals
(in that order) respecting the $D_{4h}$ point symmetry 
around the $X_3$ point.
Our model includes a pair of TNPs in contrast to the previous study focusing
on a single TNP\cite{Chang2017}. By applying symmetry
constraints and keeping terms up to quadratic order, a $\KP$ model for a pair 
of TNPs is obtained as
\begin{align}
\label{eq:hmodel-a}
\mathcal{H}\left(\bm{k}\right) =
\left(
\begin{tabular}{ccc}
    $h_{11}\left(\bm{k}\right)$ & $b_2k_xk_y$ & $c_2k_y$ \\
    $b_2k_xk_y$ & $h_{22}\left(\bm{k}\right)$ & $c_2k_x$ \\
    $c_2k_y$ & $c_2k_x$ & $h_{33}\left(\bm{k}\right)$
\end{tabular}
\right),
\end{align}
where $\bm{k}$ is a relative wave vector from the $X_3$ point and the
diagonal terms are
\begin{align}
\label{eq:hmodel-b}
    h_{11}\left(\bm{k}\right) = a_0 + a_1 k_x^2 + a_2 k_y^2 + a_3 k_z^2\,, \nonumber\\
    h_{22}\left(\bm{k}\right) = a_0 + a_2 k_x^2 + a_1 k_y^2 + a_3 k_z^2\,, \nonumber\\
    h_{33}\left(\bm{k}\right) = d_0 + d_1\left( k_x^2 + k_y^2\right) + d_3 k_z^2\,.
\end{align}

\begin{figure*} [t]
\begin{center}
\includegraphics[width=17.8cm]{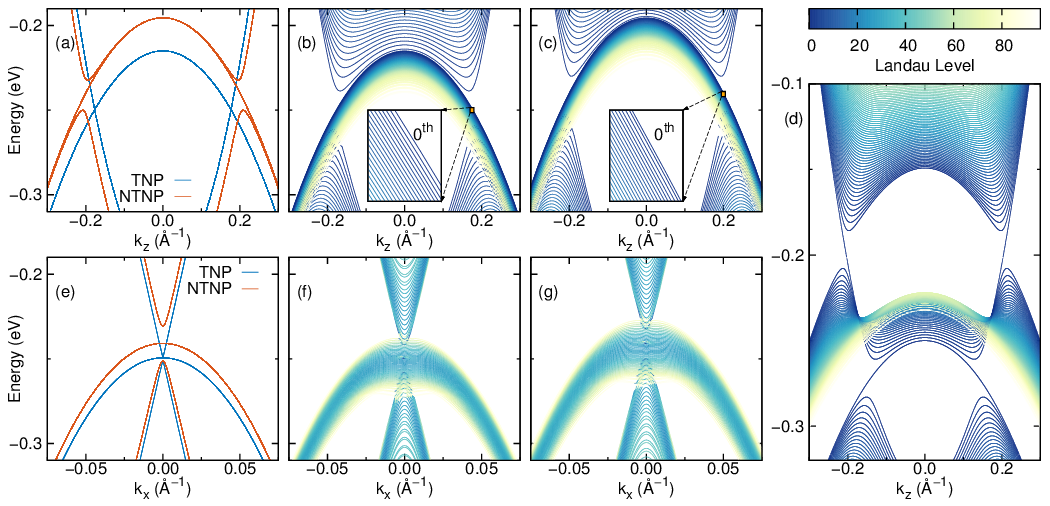}
\end{center}
\caption{Landau level spectra of GdN calculated with the $\KP$ model. (a)
and (e) Band structure calculated with the $\KP$ model along $k_z$ and
$k_x$ directions, respectively. 
(b) and (c) Landau level spectra along $k_z$ axis
for triple-nodal-point and nearly-triple-nodal-point phases, respectively.
(d) Landau-level spectra along the $k_z$
axis for the Weyl semimetal phase. (f) and (g) Landau-level spectra along
the $k_x$
direction for the triple-nodal-point and nearly-triple-nodal-point phases, 
respectively. Dark blue color in (b-d) and (f-g)
indicates low-index Lanew dau levels [see color bar in (d)].}
\label{fig:10landau}
\end{figure*}

Figures~\ref{fig:10landau}(a) and (e) show the band structure calculated
with parameters chosen to resemble the first-principles results.
Assuming an external magnetic field along the
[001] direction, the $k_x$ and $k_y$ terms are replaced by Landau-level
ladder operators according to
\begin{align}
    \pi_{x}=k_{x}-eA_{x} \rightarrow \frac{1}{\sqrt{2}l_B}\left(\hat{\textit{a}}^{+}+\hat{\textit{a}}\right),\nonumber\\
    \pi_{y}=k_{y}-eA_{y} \rightarrow \frac{1}{i\sqrt{2}l_B}\left(\hat{\textit{a}}^{+}-\hat{\textit{a}}\right),
\end{align}
where $A_i$ is the vector potential, 
$l_B = \sqrt{\hbar/eB}$ is the magnetic length, and
$\hat{\textit{a}}\left|n\right>=\sqrt{n}\left|n-1\right>$, and
$\hat{\textit{a}}^+\left|n\right>=\sqrt{n+1}\left|n+1\right>$ are the
lowering and raising operators acting on the
$n^{\textrm{th}}$ Landau level.

The Landau level spectra are then calculated with a finite number of
Landau levels in the basis. The result for the TNP phase is shown in
Fig.~\ref{fig:10landau}(b). The zeroth Landau level, the outermost of
the parabolic ones, connects valence and conduction Landau bands. 
In contrast to Weyl semimetals, additional Landau levels appear 
in the vicinity of the zeroth Landau level with a gradual shift to lower 
energy, forming a dense parabolic spectrum~\cite{Chang2017}. 
The small downward shift between subsequent Laundau levels
is determined by the negative dispersion of band 6 shown in 
Fig.~\ref{fig:1bulk_wo_soc}(c-e,h), whose parabolic dispersion 
along the direction normal to the applied magnetic field
allows higher Landau levels at lower energy.
This behavior is consistent with the results of a previous study~\cite{Chang2017}
after taking into account that the dispersion was positive there, so that
the Landau levels shifted upwards instead.

In the case of the NTNP phase, the SOC together with the
[100]-oriented magnetization lowers the symmetry considerably, but
$C_{2}^{[100]}$ and $\sigma_{[100]}$ symmetries survive.
The result is that the modified symmetry allows new terms in the
$\KP$ model of the form $k_yk_z$ in all the diagonal terms, $k_xk_z$
in $h_{12}$, and $k_z$ in $h_{13}$.%
\footnote{The reduced symmetry also implies that some coefficients
  that were identical in Eqs.~(\ref{eq:hmodel-a}-\ref{eq:hmodel-b})
  can become different.  For example, $h_{11}(k,0,0)=h_{22}(0,k,0)$ is
  no longer enforced without $C_{4}^{[001]}$ symmetry. However,
  such changes are expected to be small and have not been
  incorporated into the model used in the calculations presented here.}
The extra term in $h_{13}$ is relevant to the mass gap of the NTNP.

The Landau level spectrum of the NTNP
phase is shown in Fig.~\ref{fig:10landau}(c).
This also exhibits a zeroth Landau level connecting
valence and conduction Landau bands, spreading downward as in the TNP
phase. Although the NTNP phase gives similar results to the TNP phase and
previous studies, this specific calculation is not realistic in the sense
that the applied orbital magnetic field (along [001]) and the spin
magnetization orientation 
(along [100]) are not parallel. If the spin magnetization 
aligns with the external field, the NTNP splits into Weyl points as
shown in Fig.~\ref{fig:10landau}(d), where the upper and lower Weyl points
exhibit opposite chirality.

Therefore, we consider a case in which both the
external field and magnetization orientation are aligned along the [100]
direction by carrying out the ladder-operator replacement on the
$k_y$ and $k_z$ terms in the $\KP$ Hamiltonian.
Figures~\ref{fig:10landau}(f) and (g) show the Landau-level spectra
along the $k_x$ direction for the TNP and NTNP phases, respectively.
In contrast to previous cases [Fig.~\ref{fig:10landau}(b-d)],
the parabolic spectrum of Landau-level curves now spreads in
both directions in energy, and the zeroth Landau level does not appear
among them, since there is no longer a band
extremum in the 2D momentum space orthogonal to the axis.
In this case we find difficulty in converging the calculation with
respect to the basis set size, so that our confidence in the
accuracy of the calculation is reduced.
Nevertheless, the similarity of the magnetotransport properties between the
NTNP and TNP phases is clear, implying that half-metallic GdN can
serve as a useful platform for investigating the properties of the
TNP phase.

%=================================================
\section{Summary}
%=================================================
%
By employing \emph{ab initio} calculations, we have investigated the
topological nature of the band crossing points in half-metallic GdN. The
emergent triple nodal points in the absence of SOC split into conventional
Weyl points when taking the weak SOC into account. Interestingly, some crossing
points on the $C_4$ rotation axis orthogonal to the magnetization direction 
remain in the nearly-degenerate triple-nodal-point state.  These
``nearly triple nodal points'' induce quantitatively similar
surface spectra and transport properties compared to those of
true triple nodal points. The transition as a function of
magnetization orientation between the nearly-triple-nodal-point and
Weyl-point phases opens promising opportunities for the manipulation
of the rich surface-state structure.

\begin{acknowledgments}
We thank N. Kioussis and P.-Y. Chang for helpful discussions.
This work was supported by NSF DMR-1408838.
\end{acknowledgments}

%\bibliography{main.bbl}
%merlin.mbs apsrev4-1.bst 2010-07-25 4.21a (PWD, AO, DPC) hacked
%Control: key (0)
%Control: author (8) initials jnrlst
%Control: editor formatted (1) identically to author
%Control: production of article title (-1) disabled
%Control: page (0) single
%Control: year (1) truncated
%Control: production of eprint (0) enabled
%

\end{document}